Thermoelectric Studies of Nanoporous Thin Films with Adjusted Pore-Edge Charges

Qing Hao,[*] Hongbo Zhao, Dongchao Xu

Department of Aerospace and Mechanical Engineering, University of Arizona

Tucson, Arizona 85721, USA

Electronic mail: qinghao@email.arizona.edu

Abstract

In recent years, nanoporous thin films have been widely studied for thermoelectric applications. High thermoelectric performance is reported for nanoporous Si films, which is attributed to the dramatically reduced lattice thermal conductivity and bulk-like electrical properties. Porous materials can also be used in gas sensing applications by engineering the surface-trapped charges on pore edges. In this work, an analytical model is developed to explore the relationship between the thermoelectric properties and pore-edge charges in a periodic two-dimensional nanoporous material. The presented model can be widely used to analyze the measured electrical properties of general nanoporous thin films and two-dimensional materials.

# I. INTRODUCTION

In recent years, nanoporous thin films have been widely studied for applications in thermoelectrics,[1-3] gas sensors,[4] and thermal management.[5-7] In general, a good thermoelectric material should have a high thermoelectric figure of merit (ZT), given as ZT = $S^2\sigma T/\kappa$, where $S$, $\sigma$, $\kappa$, and $T$ represent Seebeck coefficient, electrical conductivity, thermal conductivity, and absolute temperature, respectively.[8] Here $\kappa$ can be further split into lattice contribution $\kappa_L$ and electronic contribution $\kappa_E$. Experimentally, it is found that periodic nanopores can largely decrease the lattice thermal conductivity $\kappa_L$ but still keep bulk-like power factor $S^2\sigma$, leading to high ZTs.[9-11] The same strategy has also been applied to two-dimensional (2D) materials such as graphene with periodic nano- or atomic-scale pores, called graphene antidot lattices (GALs).[12] For ~1 nm porous patterns, an electronic band gap can be possibly opened in semimetal graphene to dramatically increase its low $S$ and thus power factor $S^2\sigma$. In theoretical predictions, the combination of a high power factor and a low $\kappa_L$ leads to ZT~1.0 at 300 K for GALs.[13]

With ~10 nm porous structure sizes, the existing studies on thin films mainly focus on the $\kappa_L$ reduction. For electron transport, defects on pore edges may trap nearby charges and form a cylindrical electric field around each pore, with the field peak on the pore edge. Such an electric field has the same origin as the electric field near grain boundaries within polycrystals[14] or the particle-host interfaces within a nanocomposite,[15] where dangling bonds on these interfaces deplete nearby charges and form an energy barrier for electron transport. Based on the Boltzmann transport equation (BTE), these electric fields have been incorporated into the electrical property predictions of a polycrystal or particle-in-a-host nanocomposite. In principle, an energy barrier mostly filters out low-energy electrons, leading to reduced $\sigma$ but a higher $S$. With a suitable barrier height, this can enhance the power factor $S^2\sigma$ and thus ZT.[16] In analogy, the cylindrical electric field around nanopores can also suppress the low-energy electron transport. However, the influence of such an electric field has not been considered in existing modeling, where a constant electron relaxation time for bulk Si is simply used to estimate electrical properties.[17]  As an extreme case, a large amount of charges can be trapped on pore edges when oxidation gas molecules are adsorbed on the pore edges of porous silicon[18] or GALs[19] for gas-sensing applications.  With a very high surface-to-volume ratio, electrons within such a nanostructure can be largely depleted and thus minimize the electrical conductivity. When the material is exposed to reducing gases, electrons trapped by the oxygen adsorbate will be released, therefore restoring $\sigma$.  For atomic-thick graphene and other two-dimensional materials, pore-edge defects are more influential, which can be observed in the improved sensitivity of GAL-based gas sensors.[19]  The pore-edge defects can not only deplete the near-pore region but also affect the electronic band structure of GALs.[20,21]  By changing the gas environment, it is possible to tune the pore-edge-trapped charges and thus the thermoelectric properties of porous silicon films and GALs even during their operations.  Such opportunities are not available for internal interfaces within a bulk material due to the challenge in exactly controlling the interfacial dangling bonds during materials synthesis.  Although high-temperature annealing can be used to change the dangling bonds on internal interfaces, such changes cannot be well controlled in practice.

In this work, general two-dimensional nanoporous structures are modeled for their in-plane electrical properties, while considering the electron scattering by the cylindrical electric field around each pore.  The model is demonstrated in representative nanoporous Si films to compute their thermoelectric properties.  It is found that increased pore-edge-trapped charges often lead to monotonically decreased power factors and thus ZTs. The proposed model can be widely used for nanoporous thin films or atomic-thick materials to analyze their electrical properties.  Chemical contamination and thus pore-edge charges are inevitable more or less in the manufacture process of these materials, which often affects their electrical properties.

## II. MODELING ELECTRON AND PHONON TRANSPORT

### A. Scattering rates for nanopores with and without surface charges

In electron transport modeling, periodically arranged nanopores with uniform surface charges are considered for a quasi-two-dimensional thin film. The same approach can also be applied to atomic-thick materials. For the pore-edge electric field and its scattering rate of charge carriers, major equations are given in this section and more detailed derivations are provided in Appendix A.

Figure 1 shows one period of the structure. The radius of the nanopore is $a$ and the period is $L$ measured as the distance between centers of the adjacent nanopores. Nanopores can be arranged on a square lattice or a hexagonal lattice. The film can be doped with $N_D$ donor atoms per unit volume, or $N_A$ acceptor atoms per unit volume. On the surface of each nanopore, trapped charges with number density $N_s$ per unit area lead to an annulus depletion region with width $d$ around each nanopore (Fig. 1). For simplicity, we assume complete depletion in the depletion region ($a < r < a + d$) and charge neutrality for $r > a + d$. The charge carrier density has an abrupt jump at $= a + d$, i.e., the edge of the depletion region. Without loss of generality, we consider the case of an n-type thin film with negatively charged nanopore surfaces. The width of the annulus depletion region is calculated as

$$d = a\left(\sqrt{1 + \frac{2N_s}{aN_D}} - 1\right). \tag{1}$$

For nanopores arranged on a square lattice, the porosity of the thin film is $\varphi = \pi a^2 / L^2$. Since electrons cannot occupy either nanopores or the depletion regions surrounding them, the effective charge porosity is larger and can be defined as $\varphi_e = \pi(a + d)^2 / L^2$.

The electric field due to the surface charge is confined in the depletion region and provides an extra scattering for electrons. The energy profile of the barrier due to the electric field can be expressed as $U(r) = U_b f_b(r)$. Here $U_b$ is the maximum barrier energy that is at the edge of a nanopore:

$$U_b = U(a) = \frac{ae^2}{2\epsilon} N_s' \left( \frac{1}{2} \ln \frac{N_s'}{aN_D} - \frac{N_s}{N_s'} \right), \tag{2}$$

where $N_s' = aN_D + 2N_s$ and $f_b(r)$ is a dimensionless factor, given as

$$f_b(r) = \left[ \ln \frac{a+d}{r} + \frac{r^2}{2(a+d)^2} - \frac{1}{2} \right] \Big/ \left[ \ln \frac{a+d}{a} + \frac{a^2}{2(a+d)^2} - \frac{1}{2} \right] \tag{3}$$

in the depletion region and zero elsewhere.

The scattering of electron with surface-charged nanopores can be treated as a combination of two independent scattering mechanisms: i) charge carrier scattering by nanopores without surface charges, and ii) the scattering only by the energy barrier due to the electric field in the depletion region. The rate of scattering with surface-charged nanopores can be calculated as the sum of two scattering rates using the Matthiessen's rule,

$$\frac{1}{\tau_{cp}} = \frac{1}{\tau_{po}} + \frac{1}{\tau_b}, \tag{4}$$

where $\tau_{cp}$, $\tau_{po}$, and $\tau_b$ are charge carrier relaxation times for scattering with charge-trapped nanopores, nanopores only, and the barrier for the electric field in the depletion region, respectively. The first of the two independent scattering mechanisms, the scattering with neutral nanopores, can be simply treated as a geometric problem. This scattering can be characterized with an effective mean free path (MFP) $\Lambda_{Pore}$. For a 2D thin film with circular pores, this $\Lambda_{Pore}$ can be obtained by the Monte Carlo ray tracing technique.[22] This technique tracks the pore-edge scattering of ballistically transporting particles within a porous structure. Within a time step $\Delta t$, the percentage of particles not scattered by pore edges ($P_{\text{unscatt}}$) is related to $\Lambda_{Pore}$ by[23]

$$\Lambda_{Pore} = \frac{v \Delta t}{\ln\left(\frac{1}{P_{\text{unscatt}}}\right)}, \tag{5}$$

where the same velocity $v$ is assigned to all particles. In practice, it is found that the same $\Lambda_{Pore}$ can also be predicted by the mean beam length (MBL) widely studied in radiation.[22] Under the optically thin limit, this MBL is computed by the solid-region volume $V$ and pore-surface area $A_{Pore}$:[24]

$$MBL = \frac{4V}{A_{Pore}} = \begin{cases} \frac{2L^2 - 2\pi a^2}{\pi a}, & \text{square lattice} \\ \frac{\sqrt{3}L^2 - 2\pi a^2}{\pi a}, & \text{hexagonal lattice} \end{cases} \tag{6}$$

for the case of 2D nanoporous thin films. For atomic-thick materials with periodic nanopores, charge carrier velocities are restricted to the in-plane direction and Eq. (6) for square lattice becomes[22]

$$MBL = 1.7 \frac{L^2 - \pi a^2}{\pi a}. \tag{7}$$

Using the MBL in Eq. (6) as $\Lambda_{Pore}$, the scattering rate for neutral nanopores can be calculated as $1/\tau_{po} = v/\Lambda_{Pore}$, in which $v$ is the velocity of a charge carrier.

The second of the two independent scattering mechanisms is related to the energy barrier due to the electric field in the depletion region. This scattering is elastic and can be characterized with $k$ and $\phi$, where $k$ is the magnitude of the 2D momentum of the electron and $\phi$ is the angle between the initial and final momenta before and after the scattering. The relaxation time $\tau_b(k)$ for this scattering is an integral that can be evaluated numerically,

$$\frac{1}{\tau_b(k)} = \frac{m_\chi^*}{2\pi\hbar^3} n_{2D} \int_0^{2\pi} d\phi \, (1 - \cos\phi) |A_{uc} M(k,\phi)|^2, \tag{8}$$

where $m_\chi^*$ is the susceptibility effective mass for charge carriers, $A_{uc}$ is the area of one unit cell, $n_{2D} = \frac{1}{A_{uc}(1-\varphi)}$ is the number of nanopores per unit area of electron transport, and $M(k,\phi)$ is the matrix element of the barrier.

## B. Other typical scattering mechanisms for electrons

Besides the electron scattering by nanopores, two more scattering mechanisms are also considered for the 2D thin films of covalent compounds such as Si. These two mechanisms are intravalley acoustic phonon scattering and ionized impurity scattering. Scattering by acoustic phonons is described with the deformation potential $E_{ac}$ for the conduction band and $E_{av}$ for the valence band. The relaxation time $\tau_{AC}$ can be calculated as[25]

$$\frac{1}{\tau_{AC}} = \frac{\pi E_a^2 k_B T}{N_{deg} K \hbar} D(E) \left[ \left( 1 - \frac{E(1 - E_{av}/E_{ac})}{E_g + 2E} \right)^2 - \frac{8}{3} \frac{E(E_g + E)}{(E_g + 2E)^2} \frac{E_{av}}{E_{ac}} \right], \qquad (9)$$

where $k_B$ is Boltzmann constant, $N_{deg}$ is the band degeneracy, $K$ is bulk modulus of Si,[26] $E_g$ is the temperature-dependent band gap.[27]  Here $E$ is carrier energy measured from conduction band minimum and valence band maximum for electrons and holes, respectively. For general nonparabolic bands, the electronic density of states (DOS) $D(E)$ is given by

$$D(E) = \frac{\sqrt{2} m_d^{*\,3/2}}{\pi^2 \hbar^3} \sqrt{E + \alpha E^2} (1 + 2\alpha E). \qquad (10)$$

The nonparabolicity $\alpha = 1/E_g$ is used to describe the nonparabolic conduction band following the Kane's model,[28-30]

$$E(1 + \alpha E) = \frac{\hbar^2}{2} \left( \frac{k_l^2}{m_{e,l}^*} + 2 \frac{k_t^2}{m_{e,t}^*} \right), \qquad (11)$$

where $k_l$ and $k_t$ are the longitudinal and transverse components of the electron wave vector, respectively. For a conduction band with ellipsoidal constant energy surfaces, the electron effective masses along the longitudinal and transverse directions are $m_{e,l}^*$ and $m_{e,t}^*$, respectively. The electron DOS effective mass is $m_d^* = N_{deg}^{2/3} \left( m_{e,l}^* m_{e,t}^{*2} \right)^{1/3}$.  The electron susceptibility effective mass is $m_\chi^* = 3\left(m_{e,l}^*{}^{-1} + 2m_{e,t}^*{}^{-1}\right)^{-1}$. For parabolic bands such as the valence band of Si, $\alpha = 0$ is used.

For ionized impurity scattering, the screen length $r_s$ of Coulomb interaction is given as

$$r_s^{-2} = \frac{2^{5/2}e^2 m^* \sqrt{k_B T}}{\pi \hbar^3 \epsilon} \int_0^\infty \left(-\frac{\partial f}{\partial z}\right) \sqrt{z} \, dz, \tag{12}$$

where $z$ is reduced charge-carrier energy $z = E/k_B T$, and $f = [1 + \exp(z - \eta)]^{-1}$ is the Fermi-Dirac distribution function with reduced Fermi energy $\eta = E_F/k_B T$. The relaxation time is given as[30]

$$\frac{1}{\tau_{II}(E)} = \frac{\pi}{8hk^4}\left(\frac{4\pi Z e^2}{\epsilon}\right)^2 D(E)\left(\ln(1 + \xi) - \frac{\xi}{1 + \xi}\right)N_D, \tag{13}$$

where $k$ is charge carrier momentum, $\xi = (2kr_s)^2$, $Z$ is the number of charges contributed by each impurity atom, and $N_D$ is the number density of ionized impurities.

## C. Electrical property calculations using the scattering rates of nanopores

The electron scattering rates for pore edges can be applied to general nanoporous thin films or comparable GALs to predict their electrical properties. In typical cases, charge carriers are scattered by acoustic phonons, ionized impurity atoms, and nanopores. Their scattering rates are discussed in Sec. II.A and II.B. The total relaxation time $\tau_C$ can be calculated using Matthiessen's rule,[8,30] given as

$$\frac{1}{\tau_C} = \frac{1}{\tau_{AC}} + \frac{1}{\tau_{II}} + \frac{1}{\tau_{po}} + \frac{1}{\tau_b}. \tag{14}$$

Once $\tau_C$ is known, all electrical properties can be predicted based on the expressions obtained by solving the electron BTE. More details are given in Appendix B for the treatment of general nonparabolic bands.

Different from a solid, $\kappa_E$ and $\sigma$ of porous structures should be further multiplied by a correction factor $F(\varphi)$ that depends on the porosity $\varphi$. For a bulk material with cubically aligned spherical pores, the Eucken's factor $F(\varphi) = (1 - \varphi)/(1 + \varphi/2)$ is often used.[31,32] However, this factor is inaccurate for 2D porous structures and ANSYS simulations are carried out to find $F(\varphi)$.[22,33] By fitting the ANSYS result, $F(\varphi) = (1 - \varphi)/(1 + \varphi)$ is identified, as a Maxwell model used for GALs[34] and proposed for nanoporous Si films.[35] When a large area of a thin film is depleted, the effective porosity $\varphi_e$ should further include the depletion region to account for the reduced number of charge carriers. In $\kappa_E$ and $\sigma$ calculations, the effective electronic porosity $\varphi_e = \pi(a + d)^2/L^2$ should be used for porosity in the Maxwell's factor. Charge carriers may still travel into the depletion region and be scattered by the electric field within this region. The influence of this pore-edge field is averaged over the whole solid volume, similar to the treatment of electron scattering by ionized impurity atoms.[30] In experiments, the carrier concentration can be measured by Hall effect measurements and its reduction from the case without pore-edge trapped charges can be used to justify this $\varphi_e$. When electron MFPs are much shorter than the feature size of the nanoporous pattern, reduced carrier concentrations can be justified from $\sigma/F(\varphi)$ that is directly compared to $\sigma$ for a solid film with the same doping level. In nanoporous Si films, such comparison indicates significant charges trapped by surface states and atomic layer deposition of $Al_2O_3$ has been used for surface passivation to preserve bulk-like $\sigma$.[10] In previous studies, the reduction of charge carriers due to interfacial trapped charges is usually neglected in the electron modeling of a bulk material.[15,16] However, this reduction must be considered for 2D porous thin films or GALs due to their large surface-to-volume ratio and the significant amount of charges trapped by adsorbed gases.

The proposed electron modeling can be easily extended to GALs and Eq. (7) should be used to modify the pristine-graphene electron MFPs. When the nanopore spacing is ~100 nm or less, ballistic electron transport is often observed for GALs on a hexagonal boron nitride substrate.[36,37] In this situation, $1/\tau_C = v/\Lambda_{Pore} + 1/\tau_b(k)$ can be used. For ultrafine nanoporous sizes (~10 nm or less), the electronic band structure is dramatically changed from pristine graphene and should be incorporated in electrical property predictions. The pore-edge chemical modification by gas molecules or other atoms also has an impact on the computed electronic band structure.[20,21]

### D. Phonon transport modeling

The calculations of the lattice thermal conductivity ($\kappa_L$) is similar to that for $\sigma$, in which the same $\Lambda_{Pore}$ and correction factor $F(\varphi)$ are used. An isotropic sine-shape phonon dispersion (Born-von Karman dispersion) is assumed with identical LA and TA branches. The sine phonon dispersion is considered more accurate than the Debye model since it correctly captures the zero phonon group velocities at the first Brillouin zone boundary.[38] The dispersion using sine function is characterized by two parameters, $\omega_{max}$ and $k_0$, the maximum of phonon angular frequency and wave vector, and is simply

$$\omega = \omega_{max} \sin\left(\frac{\pi}{2}\frac{k}{k_0}\right). \tag{15}$$

Here $k_0$ and thus the equivalent atomic distance $a_D$ is computed as[8]

$$k_0 = \frac{\pi}{a_D} = (6\pi^2 N)^{1/3}, \tag{16}$$

where $N$ is the number of primitive cells per unit volume, i.e., number of atoms per unit volume divided by number of atoms per primitive cell.

Here $\kappa_L$ is given by the Holland's model[39] that is simply the kinetic relationship:

$$\kappa_L = \frac{1}{3}F(\varphi)\sum_{i=1}^{3}\int_0^{\omega_{\max}} c_i(\omega)v_{g,i}(\omega)\Lambda_i(\omega)d\omega, \tag{17}$$

where $i$ goes through all acoustic branches, $c_i(\omega)$ is spectral volumetric phonon specific heat, $v_{g,i}(\omega)$ is phonon group velocity, and $\Lambda_i(\omega)$ is the modified phonon MFP. The contribution of optical phonons to $\kappa_L$ is neglected because of their small group velocities. Using the Matthiessen's rule, $\Lambda(\omega)$ is modified from bulk phonon MFP $\Lambda_{\text{bulk}}(\omega)$ by

$$\frac{1}{\Lambda(\omega)} = \frac{1}{\Lambda_{\text{bulk}}(\omega)} + \frac{1}{\Lambda_{pore}}. \tag{18}$$

In a heavily doped bulk material, phonons are scattered by other phonons through the momentum-conserved normal process and momentum-unconserved Umklapp process, or by charge carriers. For materials such as Si, the Umklapp process dominates phonon-phonon scattering, and the relaxation time is given by[38,40]

$$\frac{1}{\tau_U} = B_1\omega^2 T \exp(-\frac{B_2}{T}). \tag{19}$$

More accurate $\kappa_L$ can be calculated using the real phonon dispersion and scattering rates obtained from first principles calculations and pump probe measurements.[41-44] However, the overall trend of spectral $\kappa_L$ contributed by phonon with different MFPs does not diverge a lot between different studies.[45] In heavily doped samples, phonons are also scattered by point defects and charge carriers. The relaxation time $\tau_I$ for phonon scattering by point defects is given by Klemens[46,47] as

$$\frac{1}{\tau_I} = A\omega^4, \tag{20}$$

where the constant $A$ includes the contribution by impurity atoms, stress/strain due to the insertion of these atoms, and unintentional defects.

The relaxation time $\tau_E$ for charge-carrier scattering of phonons is given by Ziman[45] as

$$\frac{1}{\tau_E(\omega)} = \frac{E_a^2 m^{*3} v_g}{4\pi \hbar^4 \rho} \frac{k_B T}{\frac{1}{2}m^* v_g^2} \times \left\{ \frac{\hbar \omega}{k_B T} - \ln\left[ \frac{1 + \exp\left[ \frac{\frac{1}{2}m^* v_g^2 - E_F}{k_B T} + \frac{\hbar^2 \omega^2}{8m^* v_g^2 k_B T} + \frac{\hbar \omega}{2k_B T} \right]}{1 + \exp\left[ \frac{\frac{1}{2}m^* v_g^2 - E_F}{k_B T} + \frac{\hbar^2 \omega^2}{8m^* v_g^2 k_B T} - \frac{\hbar \omega}{2k_B T} \right]} \right] \right\}, \quad (21)$$

where $E_a$, $m^*$, $v_g$, $\rho$, and $E_F$ are acoustic deformation potential, DOS effective mass, averaged phonon group velocity, density, and Fermi energy, respectively. Based on previous studies on nanoporous thin films,[22,35,48] pore edges are assumed to scatter phonons diffusively so that $\Lambda_{pore}$ in Eq. (6) can be used as a good approximation. When the dimension of nanoporous structure becomes comparable to the dominant phonon wavelength (~1 nm at 300 K[8]), phononic effects due to coherent phonon transport should be further considered. Such effects reduce the phonon group velocities and therefore $\kappa_L$. In comparable superlattices with well-defined periodic interfaces, phonon coherence becomes critical for sub-10 nm periods at 300 K, whereas diffusive phonon scattering by interfaces is dominant for larger periods.[49,50]

## III. RESULT AND DISCUSSION

### A. Model verification

A simple situation is first considered, where the nanopores have no surface charges. To compare with experimental results, power factors of p-type Si nanoporous thin films at 300 K are calculated and their doping dependence are shown in Fig. 2(a). The computed four curves are for four different geometric configurations, denoted by (a/L) for the radius $a$ and pitch $L$ of hexagonally aligned nanopores. Both $a$ and $L$ are in nanometers. Black square shows a result from an early measurement[10] and the three circles are from more recent measurements by the same research group.[11] The more recent experiment shows much lower power factors due to the use of ion implantation for doping, instead of diffusion doping in the early work. The hole mobility is much lower due to defects introduced by ion implantation.

To compare with a solid film, the obtained power factor is divided by $F(\varphi) = (1 - \varphi)/(1 + \varphi)$ to remove the porosity influence. For the (16/55), (18.5/60), and (18/60) configurations, the equivalent power factors peak at around 20 $\mu$W/cm·K$^2$, which is comparable to a power factor around 16 $\mu$W/cm·K$^2$ for the state-of-the-art bulk SiGe alloys[51] or nanostructured SiGe thin films without optimized carrier concentrations.[52]

The doping dependence of power factor is also calculated for n-type nanoporous 2D Si thin films at 1200 K as shown in Fig. 2(b). The power factor increases with higher doping, which agrees with experimental results. A high doping of $3 \times 10^{20}$ cm$^{-3}$ is thus used in the following calculation. Similarly high doping levels have been used in previous studies of nanograined bulk Si.[53,54]

For phonon transport, detailed comparison between theoretical modeling and experimental data can be found in our separated work.[22] It should be noted that experimental $\kappa$ data on fine nanoporous films are often lower than the theoretical predictions. For sub-100 nm or larger patterns, phononic effects are less important because coherent phonon transport only affects some long-wavelength phonons and the overall impact on $\kappa_L$ is very limited.[48,55] One reasonable explanation attributes the low $\kappa$ to the amorphous pore edges introduced by the nanofabrication process,[56] which effectively expands the pore diameter.[15] Since amorphous pore edges also block electron transport, the pore diameter in the proposed models can be viewed as the effective pore diameter. Although it is hard for electrons and phonons to pass through the amorphous pore edges, the surface defects within these edges can increase the charges trapped on pore edges.

**B. Thermoelectric properties of a 2D nanoporous thin film**

The developed electron transport model is used to predict the thermoelectric properties of an n-type 2D nanoporous Si thin film that has $a$ = 18 nm and $L$ = 60 nm. The nanopores are arranged on a hexagonal lattice. The structure resembles those for measured p-type thin films.[10,11] For electron modeling, a six-fold degenerate nonparabolic ellipsoidal conduction band and a parabolic spherical valence band are considered. For heavily doped n-type Si ($N_D$ >1.0×10$^{18}$ cm$^{-3}$), shallow impurity levels within the band gap start to merge with the conduction band so that the dopants are always completely ionized.[57] In this case, $Z$ =1 in Eq. (13) and the carrier concentration outside the depletion region is equal to $N_D$. The details of electrical property calculations are documented in Appendix B. For $\kappa_L$ calculations, the phonon dispersion along the (001) direction is assumed for all directions of Si. All major parameters used for electron and phonon modeling are listed in Table I.

With $N_D$ as 3×10$^{20}$ cm$^{-3}$, the thermoelectric properties are computed as a function of the absolute temperature $T$, whereas electrical properties further depend on the barrier height $U_b$. In Fig. 3, the width of the depletion region ($d$) and the charge density $N_s$ on pore surfaces are plotted as a function of the energy barrier height $U_b$ around pore edges. Both $d$ and $N_s$ increase monotonically with increased $U_b$. Here $U_b$ is simply chosen as the parameter for modeling, while $d$ and $N_s$ can be determined by $U_b$. The upper limit of $U_b$ is determined by two factors. One factor is the maximum surface charge number density. For Si, this value is restricted by the maximum atomic number density of a Si surface, which determines the number of dangling bonds for gas adsorption. For the (001) plane, there are two Si atoms per square unit cell with the lattice constant of 0.543 nm, so the Si number density for the (001) plane is ~6.78×10$^{14}$ cm$^{-2}$. In practice, this value is a few orders of magnitudes larger than the actual number density of gas molecules adsorbed onto the pore boundaries. The other factor is the maximum size of the depletion region for the proposed electron modeling, which assumes isolated depletion regions for individual pores, i.e., $a+d < L/2$. When the neighboring depletion regions overlap, the nanoporous thin film becomes an insulator. For the current geometry, $d$ should be less than 12 nm.

Figures 4(a-f) show the temperature dependence of various thermoelectric properties as calculated for the nanoporous Si film from 300 to 1200 K. Six curves are presented on each plot representing different maximum barriers ($U_b$ = 0.05, 0.2, 0.5, 1, 2, and 5 eV). Figure 4(a) shows the temperature-dependent Fermi energy $E_f$ referred to the conduction-band edge of regions outside the depletion region, which is the same for all cases due to the same doping level outside the depletion region. Because of the existence of the depletion region, the average electron density across the whole nanoporous system is smaller than the electron density outside the depletion region, i.e., the doping level $N_D$. The temperature-dependent electrical conductivity is shown in Fig. 4(b). The three curves for $U_b \leq 0.5$ eV are very close to each other, indicating that for small $U_b$ the influence of the depletion electric field is small. According to the Fermi-Dirac distribution, the probability of an electron having a kinetic energy of $E_f + 3k_BT$ is less than 5% when $E_f$ is above conduction minimum, so $E_f + 3k_BT$ can be used as the upper limit of the electron energy. In the computed cases, $E_f + 3k_BT$ can be as large as 0.5 eV so that there is some portion of electrons with energy higher than $U_b$. When $U_b > 1$ eV, the maximum barrier is greater than the energy of almost all electrons, and electrons are strongly reflected by the pore-edge electric field. As the result, the electrical conductivity drops fast as $U_b$ increases. The electronic thermal conductivity shown in Fig. 4(c) decreases with increased $U_b$ and then

increases above a threshold temperature. Known as bipolar conduction, this effect is due to the contribution of activated minority carriers at high temperatures.[8] The absolute value of Seebeck coefficient $S$ increases with increased $U_b$ as shown in Fig. 4(d), and contributing to the power factor through $S^2$. However, the decrement in the electrical conductivity dominates, which is due to enhanced electron scattering and reduced electron number. This leads to reduced powers factor in Fig. 4(e) for increased $U_b$. Finally, ZT decreases with increased $U_b$, as show in Fig. 4(f).

The barrier-height dependency of the thermoelectric properties of Si thin film are further shown in Fig. 5. In Fig. 5(a), the absolute value of Seebeck coefficient increases with increased barrier height $U_b$ at four representative temperatures. This shows the effect of filtering out the low-energy charge carriers, which is known for similar energy filtering due to the potential barrier of a grain boundary.[16] Figure 5(b) shows fast decay of the electrical conductivity as $U_b$ increases. There are two effects suppressing the electrical conductivity. One is the scattering of charge carriers by the depletion-region electric field. The variation of this effect is shown when $U_b < 2$ eV. The other effect is the reduction of the free charge carriers, as a large amount of charge carriers are electrons. The carrier number reduction is remarkable when $U_b \sim 5$ eV, as all four curves for different temperatures collapse into one in Fig. 5(b). It should be noted that the employed $U_b$ still keeps all depletion regions isolated. For even larger $U_b$, it is possible that depletion region can overlap and the material turns into an insulator, which can be the case for gas sensors with full depletion of the material.[4] The influence of $U_b$ on the total thermal conductivity is through the electronic thermal conductivity $\kappa_E$. However, this influence is negligible as $\kappa_L$ still dominates in Si thin films, which is indicated by Fig. 5(c). Although $S$ and $\sigma$ show different $U_b$ dependency, the overall effect of $U_b$ is to suppress the power factor as shown in Fig. 5(d). Since $U_b$ only slightly reduce $\kappa_E$ and thus $\kappa$, the overall ZT monotonically decreases as the pore-edge electric field increases.

## IV. SUMMARY

For 2D nanoporous thin films and comparable GALs, we have developed an analytical model to include the effect of the electric field around nanopores with trapped surface charges. The model shows that the scattering rate is higher for low-energy electrons, therefore suppressing their contributions to transport properties. Although the Seebeck coefficient is enhanced as the barrier height increases, the decrement of the electrical conductivity overshadows the Seebeck enhancement.

## V. ACKNOWLEDGEMENTS

We acknowledge the support from the U.S. Air Force Office of Scientific Research (award number FA9550-16-1-0025). An allocation of computer time from the UA Research Computing High Performance Computing (HPC) and High Throughput Computing (HTC) at the University of Arizona is gratefully acknowledged.

## APPENDIX A. DERIVATION OF THE SCATTERING RATE DUE TO PORE-SURFACE CHARGES

For simplicity, an n-type porous Si thin film is considered, with doping level $N_D$ and its nanopore surfaces negatively charged with number density $N_s$. The thickness of the thin film is $h$, which is only used for derivation but does not appear in the final expression of the scattering rate as our discussion is for 2D system. The total charge on the cylindrical surface of a nanopore is $Q_0 = -2\pi a h N_s e$, and the total charge enclosed by a cylinder with radius $r$ can be expressed as

$$Q_{\text{enc}}(r) = \begin{cases} 0\,, & r < a \\ Q_0, & r = a \\ Q_0\left[1 - \frac{r^2-a^2}{(a+d)^2-a^2}\right], & a < r < a+d \\ 0\,, & r \geq a+d \end{cases}. \tag{A1}$$

Using charge neutrality at $r = a+d$, i.e., $2\pi a h N_s = \pi[(a+d)^2 - a^2]h N_D$, we can calculate the width of the depletion region

$$d = a\left(\sqrt{1 + \frac{2N_s}{aN_D}} - 1\right). \tag{A2}$$

Applying Gauss's Law, the magnitude of the electric field (**F**) in the depletion region ($a < r < a+d$) is

$$F(r) = -\frac{N_s e}{\epsilon} \cdot \frac{a}{r} \cdot \left(1 - \frac{r^2-a^2}{(a+d)^2-a^2}\right), \tag{A3}$$

where the minus sign comes from the negative surface charge and means that the field is in negative radial direction, $\epsilon$ is the dielectric constant of Si. Setting electric potential at infinity to zero, the electric potential in the depletion region is

$$\Phi(r) = \int_r^{a+d} F(r')dr' = -\Phi_0\left[\ln\frac{a+d}{r} + \frac{r^2}{2(a+d)^2} - \frac{1}{2}\right], \tag{A4}$$

where

$$\Phi_0 = \frac{aN_s e}{\epsilon}\frac{(a+d)^2}{(a+d)^2-a^2} = \frac{ae}{2\epsilon}(aN_D + 2N_s). \tag{A5}$$

We can also get the electric potential at pore edge

$$\Phi(a) = -\Phi_b = -\Phi_0\left(\ln\frac{a+d}{a} - \frac{a^2}{2(a+d)^2} + \frac{1}{2}\right). \tag{A6}$$

Using notation $N_s' = aN_D + 2N_s$, the magnitude of pore-edge electric potential can be simplified as

$$\Phi_b = \frac{ae}{2\epsilon} N_s' \left( \frac{1}{2} \ln \frac{N_s'}{aN_D} - \frac{N_s}{N_s'} \right). \tag{A7}$$

The electric potential in depletion region provides the barrier for electron's motion. The barrier energy is $U(r) = -e\Phi(r)$ and it is peaked at the pore edge with $U_b = e\Phi_b$.

Setting 2D $|\mathbf{k}\rangle$ and $|\mathbf{k}'\rangle$ as the initial and final electronic states, respectively, the matrix element of the barrier energy is

$$M_{\mathbf{kk'}} = \langle \mathbf{k}|U(r)|\mathbf{k}'\rangle = \frac{1}{A} \iint e^{-i(\mathbf{k}'-\mathbf{k})\cdot\mathbf{r}} e\Phi(r) d^2\mathbf{r}, \tag{A8}$$

where $A = L^2$ is the area of the unit cell as shown in Fig. 1. Let $\phi$ be the angle between $\mathbf{k}$ and $\mathbf{k}'$, $\theta$ be the angle between $\mathbf{q} = \mathbf{k}' - \mathbf{k}$ and $\mathbf{r}$. Considering elastic scattering, we have $(\mathbf{k}' - \mathbf{k}) \cdot \mathbf{r} = 2k \sin\frac{\phi}{2} \cdot r \cos\theta = qr\cos\theta$. Using the identity involving Bessel function of the first kind $J_n(x)$,

$$\int_0^{2\pi} e^{iqr\cos\theta} d\theta = 2\pi J_0(qr), \tag{A9}$$

the matrix element can be rewritten as

$$M_{\mathbf{kk'}} = \frac{\pi(a+d)^2 e\Phi_0}{A} \int_\gamma^1 (-x + x^3 - 2x \ln x) J_0[q(a+d)x] dx, \tag{A10}$$

where $x = \frac{r}{a+d}$ and $\gamma = \frac{a}{a+d}$. After integrating Eq. (A10), we get the matrix element

$$M_{\mathbf{kk'}} = \frac{M_{\mathbf{kk'}}'}{A} = \frac{1}{A}\pi(a+d)^2 e\Phi_0[I(1) - I(\gamma)], \tag{A11}$$

where

$$I(x) = -\frac{1+2\ln x}{p} x J_1(px) + \frac{2}{p^2}\left(1 - J_0(px) + x^2 J_2(px)\right) - \frac{1}{p} x^3 J_3(px) \tag{A12}$$

with $p = (a+d)q$.

Using Fermi's golden rule, the rate of electron scattering with the barrier in depletion region is

$$W(\mathbf{k}, \mathbf{k}') = \frac{2\pi}{\hbar} \left| \frac{1}{A} M_{\mathbf{kk'}}' \right|^2 \delta\big(E(\mathbf{k}) - E(\mathbf{k}')\big) \tag{A13}$$

and the scattering rate due to the electric field barrier is

$$\frac{1}{\tau_b(\mathbf{k})} = \frac{A}{(2\pi)^2} \iint (1 - \cos\phi) W(\mathbf{k}, \mathbf{k}') d^2\mathbf{k}'. \tag{A14}$$

Changing the integration in 2D polar coordinates, Eq. (A14) is

$$\frac{1}{\tau_b(k)} = \frac{1}{2\pi\hbar A} \int_0^\infty k \, dk \int_0^{2\pi} d\phi (1 - \cos\phi) |M'_{\mathbf{k},\mathbf{k}'}|^2 \delta\big(E(k) - E(k')\big). \tag{A15}$$

Using $k \, dk = \frac{m}{\hbar^2} dE$, after integrating out the delta function, we have the expression for the relaxation time

$$\frac{1}{\tau_b(k)} = \frac{m_\chi^*}{2\pi\hbar^3} n_{2D} \int_0^{2\pi} d\phi \, (1 - \cos\phi) |M'_{\mathbf{k}\mathbf{k}'}|^2, \tag{A16}$$

where $m_\chi^*$ is the mass of electron, $n_{2D} = \frac{1}{A(1-\varphi)}$ is the number of nanopores per unit area of electron transport, and momentum $k$ satisfies $\frac{\hbar^2 k^2}{2m} = E(k)$. From Eqs. (A11) and (A12), matrix element is just a function of $q = 2k \sin\frac{\phi}{2}$. For a given electron energy $E$, we can get its momentum $k$, and Eq. (A16) can be easily evaluated numerically.

## APPENDIX B. ELECTRICAL PROPERTY CALCULATIONS

For a solid material, the electrical conductivity for each band is given as[8]

$$\sigma_i = \frac{e^2 (2m_d^* k_B T)^{3/2}}{3 m_\chi^* \pi^2 \hbar^3} \int_0^\infty \left(-\frac{\partial f}{\partial z}\right) \tau(z) \frac{(z + \beta z^2)^{3/2}}{1 + 2\beta z} dz, \tag{B1}$$

where $z$ is reduced charge-carrier energy $z = E/k_B T$, $k_B$ is Boltzmann constant, $\beta = k_B T/E_g$, and $f = [1 + \exp(z - \eta)]^{-1}$ is the Fermi-Dirac distribution function with reduced Fermi energy $\eta = E_F/k_B T$. Here $E$ refers to the band extrema, i.e., the conduction band minimum for electrons and the valence band maximum for holes. The subscript $i$ can be either $e$ for conduction band or $h$ for valence band. The total electrical conductivity is simply $\sigma = \sigma_e + \sigma_h$.

The Seebeck coefficient for each band is given as[8]

$$S_i = \mp \frac{k_B}{e} \frac{\int_0^\infty \left(-\frac{\partial f}{\partial z}\right)\tau(z)(z-\eta)\frac{(z+\beta z^2)^{3/2}}{1+2\beta z}dz}{\int_0^\infty \left(-\frac{\partial f}{\partial z}\right)\tau(z)\frac{(z+\beta z^2)^{3/2}}{1+2\beta z}dz}, \qquad (B2)$$

with "$-$" for conduction band and "$+$" for valence band. The overall Seebeck coefficient is obtained by weighting the contribution of each band based on its normalized electrical conductivity,[8] i.e.,

$$S = \frac{S_e\sigma_e + S_h\sigma_h}{\sigma_e + \sigma_h}. \qquad (B3)$$

The electronic thermal conductivity $\kappa_E$ is calculated using Wiedemann-Franz law by

$$\kappa_E = L_0\sigma T, \qquad (B4)$$

where the Lorenz number $L_0$ is the sum of three terms: the contribution from conduction band $L_e$, from valence band $L_h$, and bipolar contribution $L_b$. The contribution from each band is given as[8]

$$L_i = \left(\frac{k_B}{e}\right)^2 \left[\frac{\int_0^\infty \left(-\frac{\partial f}{\partial z}\right)\tau(z)z^2\frac{(z+\beta z^2)^{3/2}}{1+2\beta z}dz}{\int_0^\infty \left(-\frac{\partial f}{\partial z}\right)\tau(z)\frac{(z+\beta z^2)^{3/2}}{1+2\beta z}dz} - \left(\frac{\int_0^\infty \left(-\frac{\partial f}{\partial z}\right)\tau(z)z\frac{(z+\beta z^2)^{3/2}}{1+2\beta z}dz}{\int_0^\infty \left(-\frac{\partial f}{\partial z}\right)\tau(z)\frac{(z+\beta z^2)^{3/2}}{1+2\beta z}dz}\right)^2\right], \qquad (B5)$$

and the bipolar Lorenz number is given as

$$L_b = \frac{\sigma_e\sigma_h}{\sigma^2}(S_e - S_h)^2. \qquad (B6)$$

The Wiedemann-Franz law and Eq. (B5) assume that charge carriers are elastically scattered by a pore edge and its associated cylindrical electric field. In this situation, these scattering mechanisms affect the thermal and electrical currents carried by electrons or holes in the same way to validate the Wiedemann-Franz law. For polycrystalline materials, breakdown of the Wiedemann-Franz law is reported when some heat carried by charge carriers can be transferred across a grain boundary via phonons though these charge carriers are reflected back by the grain boundary. This leads to a higher Lorentz number than predictions by Eq. (B5).[58-60] These effects are not considered for pore-edge scattering of charge carriers in this work.

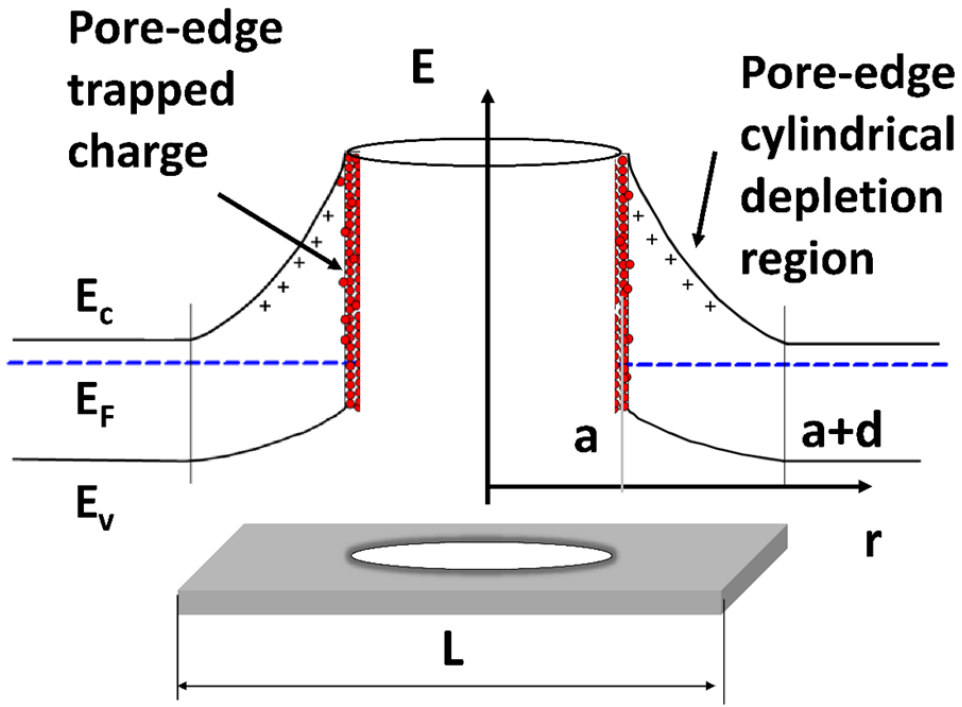

Fig. 1. Schematic diagram of one unit cell of the nanoporous film. Radius of the nanopore is *a*, width of the depletion region is *d*. Nanopores are aligned on a square lattice, with lattice constant *L*. $E_c$ and $E_v$ indicate the edge of conduction and valence bands, respectively. $E_F$ is the Fermi energy.

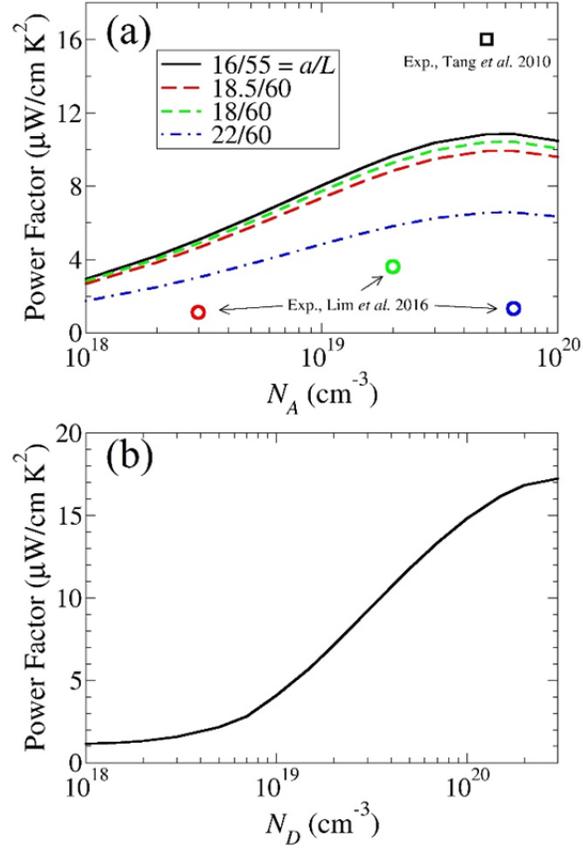

Fig. 2. Doping dependence of power factors for (a) p-type and (b) n-type nanoporous Si thin films at 300 K and 1200 K, respectively. The nanopores considered here have no surface charges. In (a), four curves are for four different pairs of pore radius and pitch. Square is for experimental data extracted from Ref. [10], and circles are for experimental data from Ref. [11]. In (b), the power factor is calculated for a film with nanopores arranged on a hexagonal lattice with $a$=18 nm and $L$=60 nm.

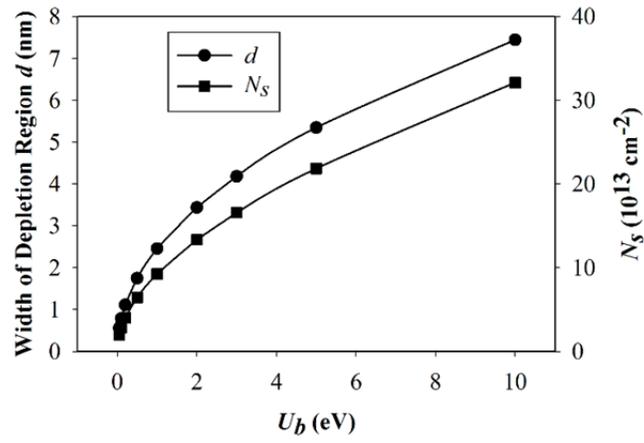

Fig. 3. The width $d$ of the depletion region (circles, left y-axis) and surface-charge number density $N_s$ (squares, right y-axis) as functions of barrier height $U_b$.

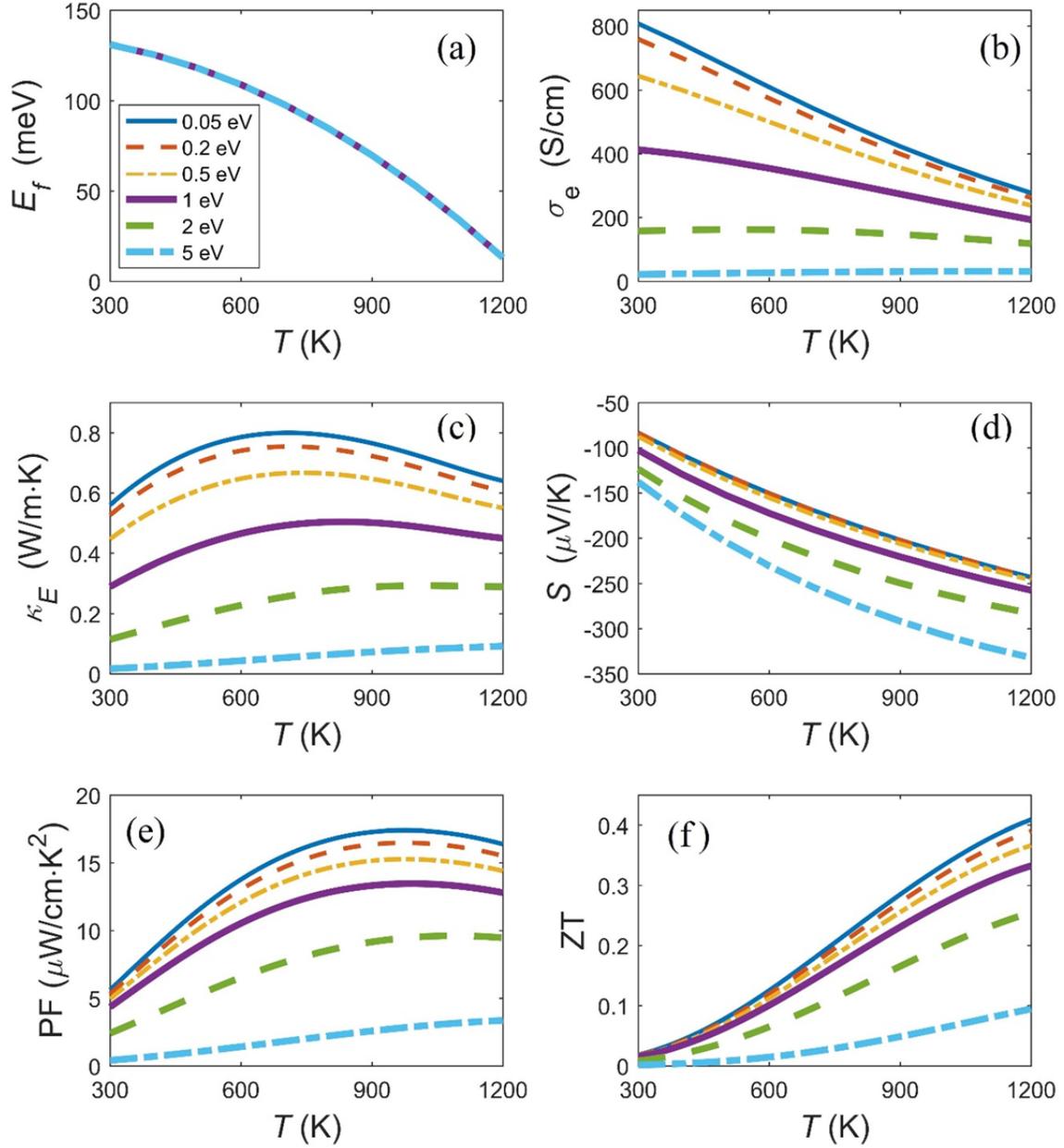

Fig. 4. (Color online) The temperature dependence of (a) Fermi energy, (b) electrical conductivity, (c) electronic thermal conductivity, (d) Seebeck coefficient, (e) power factor $S^2\sigma$, and (f) ZT as calculated for the nanoporous Si film. Legend in part (a) also applies to other parts. Different color indicates different barrier height $U_b$, which ranges from 0.05 eV to 5 eV.

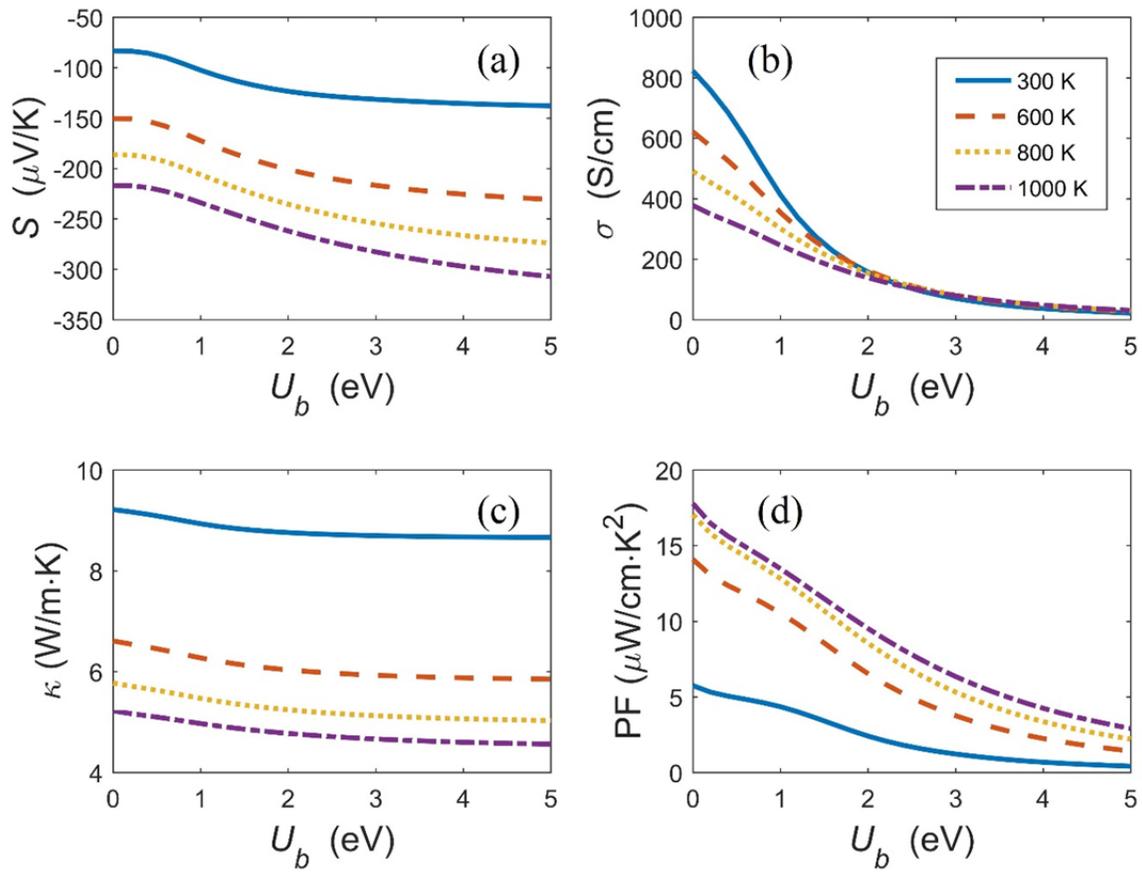

Fig. 5. (color online) The barrier height dependency of (a) Seebeck coefficient, (b) electrical conductivity, (c) thermal conductivity, and (d) power factor of the nanoporous Si thin film at four different temperatures. Legend in part (b) also applies to other parts. Different colors and line types indicate different temperatures.

Table I. Parameters used for solving the BTE for Si nanoporous film. The $m_e$ in table means the mass of electron.

| Parameter | Symbol | Value |
|---|---|---|
| Electron effective mass, longitudinal | $m_{e,l}^*$ | $0.92\ m_e$ |
| Electron effective mass, transverse | $m_{e,t}^*$ | $0.19\ m_e$ |
| Conduction band degeneracy | $N_{deg}$ | 6 |
| Hole effective mass | $m_h^*$ | $1.2\ m_e$ |
| Valence band degeneracy | $N_{deg}$ | 1 |
| Band gap | $E_g$ | $1.17 - 4.73 \times 10^{-4}\,T^2/(T+636)$ eV [a] |
| Debye temperature | $\Theta$ | 640 K |
| Bulk modulus | $K$ | $9.8 \times 10^{10}$ N/m$^2$ [b] |
| Dielectric constant | $\epsilon$ | $20\ \epsilon_0$ |
| Acoustic phonon deformation potential, electron | $E_{ac}$ | 9 eV |
| Acoustic phonon deformation potential, hole | $E_{av}$ | 3.8 eV |
| n-type doping level | $N_D$ | $3 \times 10^{20}$ cm$^{-3}$ |
| Phonon point defect scattering parameter | $A$ | $1.175 \times 10^{-44}$ S$^3$ [c] |
| Umklapp scattering parameter | $B_1$ | $1.7 \times 10^{-19}$ s/K [d] |
| Umklapp scattering parameter | $B_2$ | 210 K [d] |

[a] Ref [27]

[b] Ref [26]

[c] Ref [47]

[d] Ref [38]